\documentclass[aps,prl,twocolumn,superscriptaddress,showpacs]{revtex4-1}

\usepackage[pdftex]{graphicx}

\usepackage{bm}

\usepackage{times}

\begin{document}

\title{Out-of-plane nesting driven spin spiral 
in ultrathin Fe/Cu(001) films}

\author{J.~Miyawaki}
\affiliation{Excitation Order Research Team, RIKEN SPring-8 Center, Sayo-cho, Hyogo, 679-5148, Japan}
\author{A.~Chainani}
\affiliation{Excitation Order Research Team, RIKEN SPring-8 Center, Sayo-cho, Hyogo, 679-5148, Japan} 
\affiliation{Coherent X-ray Optics Laboratory, RIKEN SPring-8 Center, Sayo-cho, Hyogo, 679-5148, Japan}
\author{Y.~Takata}
\affiliation{Excitation Order Research Team, RIKEN SPring-8 Center, Sayo-cho, Hyogo, 679-5148, Japan} 
\affiliation{Coherent X-ray Optics Laboratory, RIKEN SPring-8 Center, Sayo-cho, Hyogo, 679-5148, Japan}
\author{M.~Mulazzi}
\affiliation{Excitation Order Research Team, RIKEN SPring-8 Center, Sayo-cho, Hyogo, 679-5148, Japan} 
\author{M.~Oura}
\affiliation{Excitation Order Research Team, RIKEN SPring-8 Center, Sayo-cho, Hyogo, 679-5148, Japan} 
\affiliation{Coherent X-ray Optics Laboratory, RIKEN SPring-8 Center, Sayo-cho, Hyogo, 679-5148, Japan}
\author{Y.~Senba}
\affiliation{JASRI/SPring-8, Sayo-cho, Hyogo, 679-5198, Japan}
\author{H.~Ohashi}
\affiliation{JASRI/SPring-8, Sayo-cho, Hyogo, 679-5198, Japan}
\author{S.~Shin}
\affiliation{Excitation Order Research Team, RIKEN SPring-8 Center, Sayo-cho, Hyogo, 679-5148, Japan} 
\affiliation{The University of Tokyo, Institute of Solid State Physics, Kashiwa, Chiba, 227-8581, Japan}


\begin{abstract}
 Epitaxial ultrathin Fe films on fcc Cu(001) exhibit a spin spiral (SS), 
 in contrast to the ferromagnetism of bulk bcc Fe. 
 We study the in-plane and out-of-plane Fermi surfaces (FSs) of the SS
 in 8 monolayer Fe/Cu(001) films using energy dependent soft x-ray
 momentum-resolved photoemission spectroscopy. 
 We show that the SS originates in nested regions confined to
 out-of-plane FSs, 
 which are drastically modified compared to in-plane FSs. 
 From precise reciprocal space maps in successive zones, 
 we obtain the associated real space compressive strain of
 1.5$\pm$0.5\% along $c$-axis.  
 An autocorrelation analysis quantifies the incommensurate ordering
 vector $\bm{q}$=(2$\pi/a$)(0,0,$\sim$0.86), favoring a SS
 and consistent with magneto-optic Kerr effect experiments. 
 The results reveal the importance of in-plane and out-of-plane FS
 mapping for ultrathin films.
\end{abstract}

\maketitle

Fe metal is delicately perched on a magneto-structural instability, with
antiferromagnetic Cr and Mn to its left, and ferromagnetic Co and Ni to
its right in the periodic table.  
In spite of its high ferromagnetic Curie temperature 
($T_\mathrm{C}$=1043~K) and large magnetic moment (=2.22~$\mu_\mathrm{B}$) in
its bulk body-centered-cubic (bcc) form, theoretical studies have
predicted face-centered-cubic (fcc) Fe  
to be nonmagnetic, ferromagnetic, antiferromagnetic or exhibit a 
spin spiral (SS: corresponding to a periodic variation
of the angle of the spins), sensitively depending on its lattice parameter
\cite{Herring1966, Wang1985, Uhl1992, Korling1996}.
The phase diagram of Fe metal is indeed very rich with, 
(1) the bcc to fcc Bain transformation at 1184~K \cite{Okatov2009}, 
(2) a hexagonal-close-packed (hcp) non-magnetic phase at high pressure
(P) and low temperature (T) which shows superconductivity
\cite{Shimizu2001}, 
while 
(3) the high-P, high-T phase was debated in terms of a hcp or
orthorhombic structure \cite{Dubrovinsky1998}. 
Also, fcc Fe nanoparticle precipitates (diameter $\sim$ 50~nm) stabilized 
in a Cu matrix shows a SS state with an
ordering vector $\bm{q}$ = (2$\pi/a$)(0.12,0,1.0) and a low-spin moment 
(=0.7~$\mu_\mathrm{B}$) per atom \cite{Tsunoda1989, Tsunoda1993}.
Remarkable progress has been achieved in understanding the
interplay of crystal structure and electronic instabilities in realizing
spin order and its influence on properties of complex materials such as
multiferroic oxides, superconducting copper-oxides \cite{Ramesh2007}
and iron-pnictides \cite{Zhao2008}, etc. 
Surprisingly, despite extensive efforts, the simpler SS
associated with the bcc-fcc ($\alpha$-$\gamma$)  
transition in elemental Fe and its momentum-resolved electronic
structure has remained a challenging unsolved problem for over 40 years
\cite{Herring1966}.

In contrast to experiments at high-T, high-P or
nanoparticles confined in a matrix, a simpler route was developed to
study structure property correlations in Fe: Fe thin films grown
epitaxially on fcc Cu(001) 
substrate
\cite{Kief1993,Ellerbrock1995,Bernhard2005,Meyerheim2005}.
Fe thin films on Cu(001) also show a complex magnetic and structural
phase diagram
\cite{Kief1993,Bernhard2005,Meyerheim2005,Li1994,Donath2009}
and it is known that
(1) Below 4 monolayers (MLs), it is ferromagnetic and has the
face-centered-tetragonal (fct) structure (Region I). 
(2) Between 5--11~ML, it has the fcc structure and a SS, with 
a top bilayer which is ferromagnetic (Region II).
The SS ordering T is $T_\mathrm{SS}{\sim}$200~K \cite{Li1994,Qian2001}. 
(3) Above 12~ML, it transforms to the bulk ferromagnetic bcc structure
(Region III). 
The SS phase of Region II is the focus in this work, and as determined
by magneto-optic Kerr effect measurements, it has a SS ordering
vector of $q_z$ = 2$\pi$/2.6$d$ or 2$\pi$/2.7$d$ 
(where $d$ is the interlayer distance of the Fe thin film)
\cite{Li1994,Qian2001}.
This corresponds to $\bm{q}$ = (2$\pi/a$)(0,0,$\sim$0.75) (where $a$ = 2$d$ is
the lattice parameter of the Fe thin films). 
The magneto-optic Kerr effect (MOKE) experiments provided the
$z$(out-of-plane)-component of the SS vector and ruled out a type-1
antiferromagnetic structure. The obtained average magnetic moment per
atom was $\sim$1.5~$\mu_\mathrm{B}$, significantly larger than for the
nanoprecipitates \cite{Li1994,Qian2001}. 
A depth resolved X-ray magnetic
circular dichroism study indicated a magnetic moment of
1.7~$\mu_\mathrm{B}$ for the SS phase, thus confirming a clear
difference compared to the nanoprecipitates \cite{Amemiya2004}. 
The possibility of a SS in fcc Fe has been 
extensively discussed from theory. The salient features for fcc Fe from
theory indicates absence of Fermi surface (FS) nesting but allows for 
SSs with ordering vectors $\bm{q}_1$ = (2$\pi/a$)(0,0,0.6) or
$\bm{q}_2$ = (2$\pi/a$)(0.5,0,1.0) \cite{Korling1996,Uhl1992}.
Both these SSs show high-spin
moment ($\sim$1.5~$\mu_\mathrm{B}$) and total energy considerations
indicate $\bm{q}_1$ is 
slightly more stable than $\bm{q}_2$. Also, a
spin density wave (SDW: corresponding to a periodic
modulation in the magnitude of collinear spins) was found to be 
extremely unfavourable compared to a non-collinear SS
\cite{Spisak2002}.
It is noted that $\bm{q}_1$ and $\bm{q}_2$ do not match with $\bm{q}$ 
of the MOKE measurements. The most 
serious limitation to date, however, is the absence of experimental
results relating the momentum ($k$) resolved electronic structure to the
SS in epitaxial ultrathin Fe/Cu(001) films.  

In order to make this connection, it is necessary to determine the
in-plane and out-of-plane FSs of ultrathin 
Fe/Cu(001) films. While in-plane FSs are routinely measured with a fixed
photon energy and angle-resolved photoemission spectroscopy (ARPES),
the specific requirement of investigating the $q_z$ component of the
SS in the 
out-of-plane FSs necessarily requires ARPES with tunable energy
\cite{Venturini2008, Kamakura2006}. While crystalline order along the
$c$-axis has been reported for 6--8~ML Fe/Cu(001) films using surface
diffraction \cite{Meyerheim2005},
the present experiments constitute the first out-of-plane FS measurements
for ultrathin films. Early study of the thickness dependent valence
band structure of Fe/Cu(001) films identified distinct changes 
at 5 ML thickness, the borderline of regions I and II
\cite{Madjoe1999}.
FS mapping carried out with a fixed energy
($h\nu$ = 75~eV), showed diffuse in-plane FSs which could not be
satisfactorily explained by generalized gradient approximation band
structure calculations of the bulk bcc or fcc structure.
The added complication of the top ferromagnetic bilayer motivated our
use of soft x-ray (SX)-ARPES as the technique of choice due to its
larger probing depth, typically 10--20~\AA.
Recent studies using SX-ARPES have reported the 3-dimensional (D)
momentum resolved electronic structure of solids
\cite{Plucinski2008,Venturini2008,Kamakura2006}. 
In a very recent work, 
it was used to obtain the 3D charge density of bulk
copper metal \cite{Mansson2008}. Hence, in comparison to low-energy ARPES,
energy-dependent SX-ARPES ($h\nu\sim$ 400--800~eV) is well-suited to
investigate the 3D FSs of the SS phase in ultrathin Fe/Cu(001) films.

\begin{figure}[htbp]
 \begin{center}
  \includegraphics[]{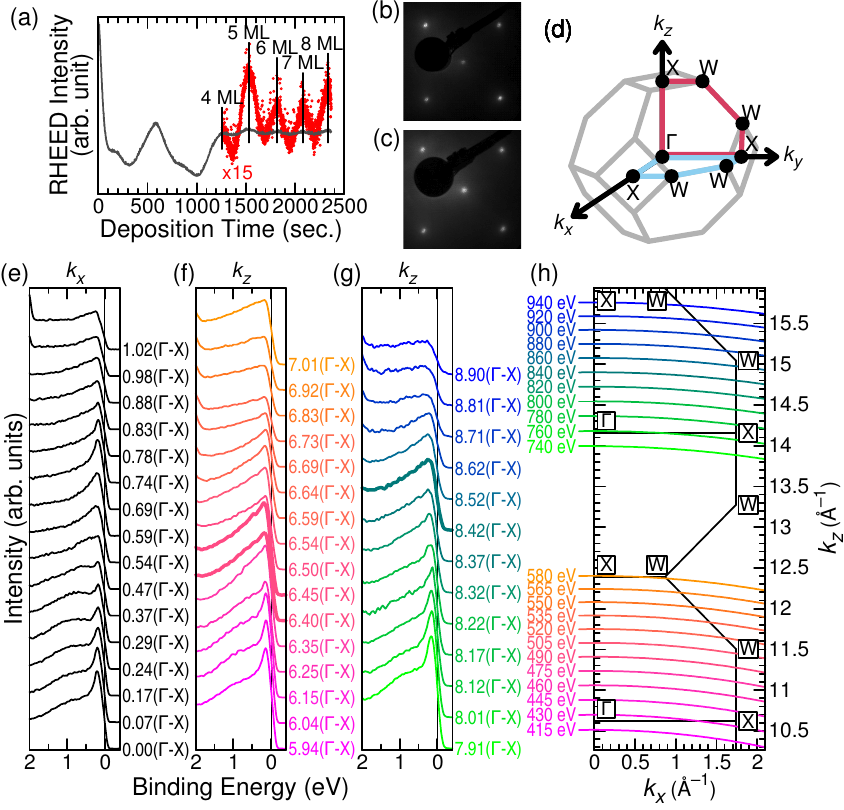}
  \caption{Characterization of the sample and determination of $k_z$.
  (a) Typical RHEED oscillations for Fe/Cu(001) grown at RT.
  (b) and (c) LEED patterns for clean Cu(001) substrate and 8~ML Fe
  thin films on Cu(001). 
  (d) The volume BZ of fcc Fe. The in-plane and out-of-plane regions
  probed in the present study are marked by blue and red lines,
  respectively. 
  (e) EDCs along in-plane $\Gamma$-X ($k_x$) measured at $h\nu$ = 430~eV. 
  (f) and (g) EDCs along out-of-plane $\Gamma$-X ($k_z$) measured using
  $h\nu$ = 415--580 and 740--940~eV.  
  (h) The calculated $k_z$ positions from the photon energy dependence
  of EDCs. 
  We used an inner potential and work function of 6 and 5~eV,
  respectively.
  The colors of the arcs corresponds to EDCs in (f) and (g).
  The EDCs plotted as thick lines correspond to momenta associated with
  the SS, as marked by arrows in Fig.~\ref{fig3}(e).}
  \label{fig1}
 \end{center}
\end{figure}

Epitaxial Fe films were grown on Cu(001)
using electron beam deposition at room temperature (RT) \cite{EPAPS}.
The thickness of the Fe thin films was controlled to 8~ML
by monitoring reflection high energy electron diffraction (RHEED)
oscillations (Fig.~\ref{fig1}(a)). 
The crystallinity and crystal orientation of the Cu(001) substrate and
the deposited Fe thin films were measured by low energy electron
diffraction (LEED) (Fig.~\ref{fig1}(b) and (c), respectively).
The RHEED oscillations and LEED patterns are in very good agreement with 
earlier work~\cite{Li1994,Qian2001,Bernhard2005},
indicating the high quality of the epitaxial ultrathin 8~ML 
Fe/Cu(001) films.

\begin{figure}[htbp]
 \begin{center}
  \includegraphics[]{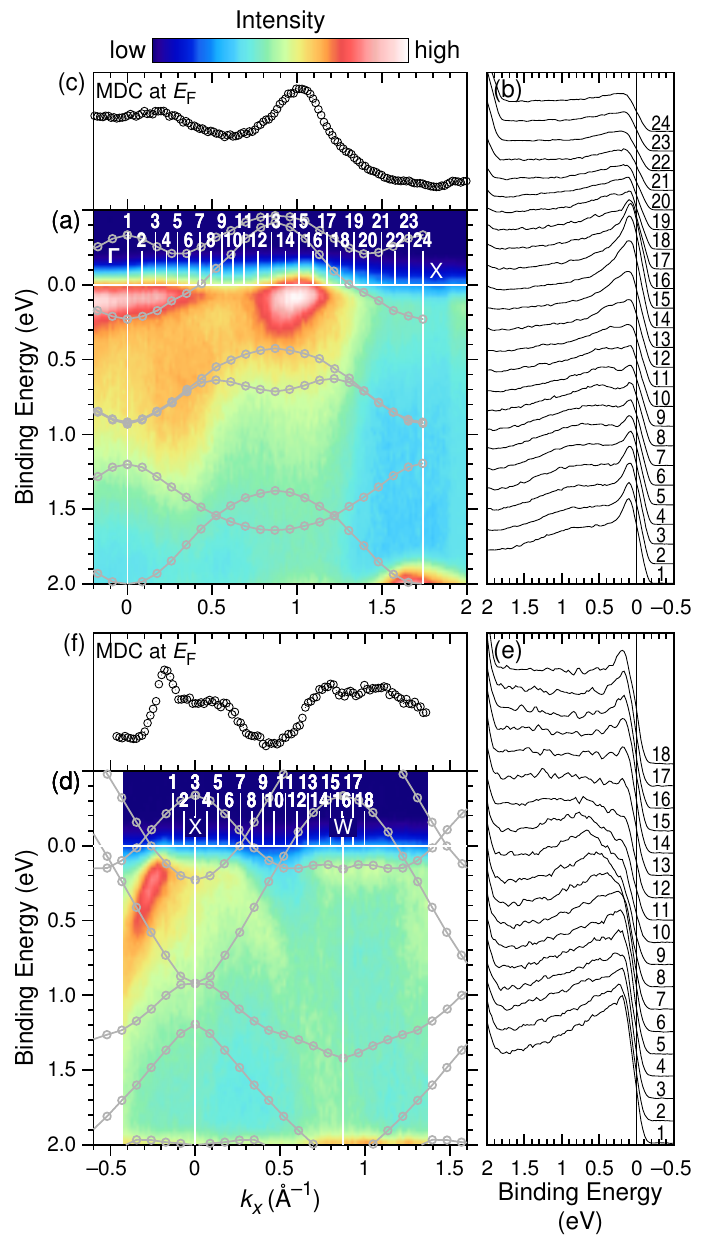}
  \caption{  ARPES results of Fe(8~ML)/Cu(001) at 50~K.
  (a) and (d) In-plane band maps measured at $h\nu$ = 430 and 580~eV. 
  The gray lines and points are the band calculation of fcc Fe
  for $\bm{q}$ = (2$\pi/a$)(0.5,0,1) \cite{Korling1996}.
  (b) and (e) EDCs at some $k_x$ points for each photon energy. 
  The numbers labeled on EDCs indicate the probed $k_x$ positions,
  as shown in (a) and (d).
  (c) and (f) MDCs at $E_\mathrm{F}$.}
  \label{fig2}
 \end{center}
\end{figure}

SX-ARPES measurements were performed with an
electron energy analyzer, 
Gammadata-Scienta SES2002, at undulator beamline BL17SU of
SPring-8 using a grazing incidence geometry ($<$5$^\circ$)
spectrometer \cite{EPAPS, Ohashi2007,Takata2004}. 
The grazing incidence spectrometer makes 
the ARPES measurements highly efficient and also ensures that the x-ray
photon momentum ($\sim$0.1$\Gamma$-X at $h\nu$ = 400~eV) imparted to the
electrons are accounted for easily \cite{Plucinski2008,Kamakura2006}.
Circularly polarized x-rays were used to avoid the 
symmetry selectivity of linearly polarized x-rays \cite{EPAPS}.
The measurements were carried out
at 50~K, which is well below $T_\mathrm{SS}{\sim}$200~K,
to minimize indirect transition losses.
The $k_x$ and $k_z$ positions in the 3D Brillouin zone (BZ)
(Fig.~\ref{fig1}(d)) were accurately determined from energy 
and momentum distribution curves (EDCs and MDCs) obtained from intensity
maps measured for incident photon energies probing two successive
BZs. 
The arcs in $k_x$-$k_z$ momentum space traced by the specific
photon energies used are shown in Fig.~\ref{fig1}(h). 
Figure~\ref{fig1}(e) shows the EDCs for the
$k_x$ scan along $\Gamma$-X at $h\nu$ = 430~eV, while
Fig.~\ref{fig1}(f) and (g) show EDCs for $\Gamma$-X along $k_z$ in
successive BZs, respectively, as marked in Fig.~\ref{fig1}(h). The
overall similarity of the EDCs, particularly at the high symmetry points
$\Gamma$ and X along $k_x$ and $k_z$ confirms the viability of 3D FS
mapping of the Fe films. 
However, since we are measuring ultrathin films with soft x-rays,
we have to be careful about contributions from the copper substrate in
the EDCs. It is well-known that copper shows very weak
intensity due to a $s$ band crossing near X point (along $\Gamma$-X and
X-W) within a binding energy of 2~eV \cite{Courths1984, EPAPS}. 
We have actually confirmed the same by measuring the clean copper
substrate ARPES spectra before the film depositions \cite{EPAPS}.
Then, as we show in the following results, the SS associated
electronic states are well separated from the copper states in energy and
$k$-space. Thus, the intensity maps within 2~eV binding energy are
dominated by the Fe spectral weight. Fig.~\ref{fig2}(a) and (b)
show the in-plane $\Gamma$-X and X-W intensity maps 
of the Fe films,
obtained using $h\nu$ = 430~eV and 580~eV,
respectively. Figure~\ref{fig2}(b)/(e) and (c)/(f) are 
the corresponding EDCs and MDCs which show clear band crossings.
Early work assigned the band crossings along $\Gamma$-X to the
$\Delta_1$ and $\Delta_5$ bands, 
as suggested by calculations for the parent high-spin fcc Fe
\cite{Mankey1993}. 
A comparison with calculated band dispersions for the
SS vectors 
obtained in the local spin density approximation (LSDA), however, does
not give a satisfactory match with the data for $\bm{q}_1$ =
(2$\pi/a$)(0,0,0.6). For $\bm{q}_2$ = (2$\pi/a$)(0.5,0,1.0), the
calculated data are overlaid in Fig.~\ref{fig2}(a) and (b)
 after shifting the energy by 0.4~eV toward Fermi edge. 
While the data show similarities with calculations, e.g.,
MDCs show band crossings matching the calculation for the X-W cut or for
the crossing at 0.35~\AA$^{-1}$ along $\Gamma$-X, there are clear
discrepancies also observed: the intense band crossing at
1.05~\AA$^{-1}$ along $\Gamma$-X and the band dispersions between 0.5
to 2~eV binding energies do not match well with calculations. 
Thus, while band dispersions and crossings can be clearly observed, 
the experimental results do not match the LSDA results for the SSs with
$\bm{q}_1$ = (2$\pi/a$)(0,0,0.6) and  
$\bm{q}_2$ = (2$\pi/a$)(0.5,0,1.0).

\begin{figure}[htbp]
 \begin{center}
  \includegraphics[]{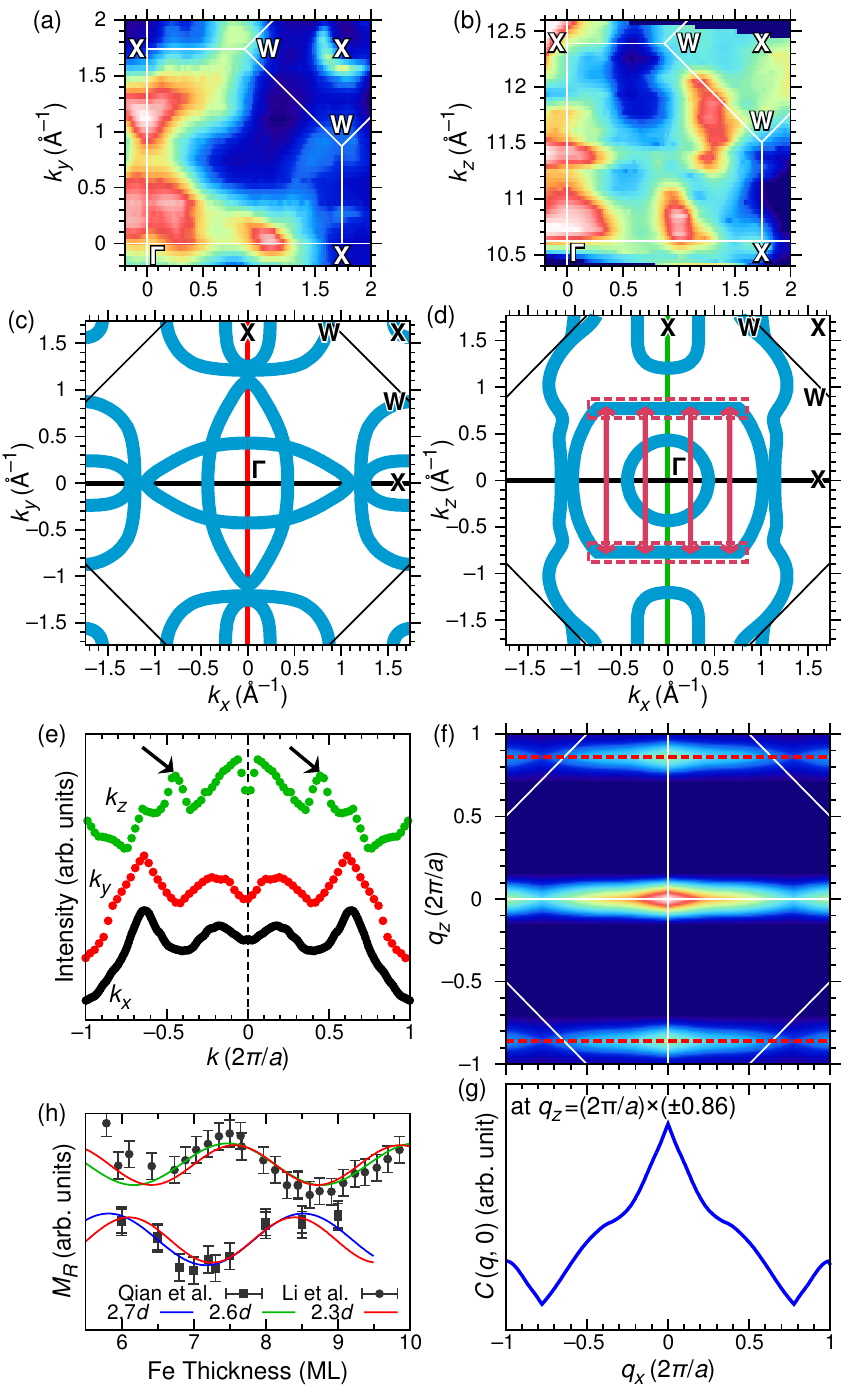}
  \caption{In-plane and out-of-plane FSs of Fe(8~ML)/Cu(001) and
  autocorrelation analysis 
  (a) and (b) In-plane and out-of-plane FSs measured
  over 1/4 of the BZ. 
  (c) and (d) In-plane and out-of-plane FS crossings
  extracted from MDCs. 
  FSs for full BZ were obtained by symmetrizing
  data from 1/4 of BZ.
  The regions indicated by the red dashed lines are the nesting parts
  and were used for the ACA.
  (e) MDCs for $\Gamma$-X along $k_x$, $k_y$ and $k_z$. 
  The $k$-cut positions are marked by black, red and green lines 
  in (c) and (d). 
  $a$=3.61 and 3.55 \AA$^{-1}$ are used for the in-plane 
  and out-of-plane lattice parameters.
  The peaks observed only in the MDC along $k_z$ are
  indicated by the arrows and the corresponding EDCs are 
  emphasized as thick lines in Fig.~\ref{fig1}(f) and (g).
  (f) AC map for the restricted regions in out-of-plane FS.
  (g) Line profile of AC map at $q_z$ = (2$\pi/a$)$\times\pm$0.86, 
  indicated by red dashed lines in (f).
  (h) The comparison of the present result and early MOKE
  measurements: SS period
  $\phi_\mathrm{SS}$ = 2.3$d$ (present result), 
  $\phi_\mathrm{SS}$ = 2.6$d$ \cite{Li1994} and 
  $\phi_\mathrm{SS}$ = 2.7$d$ \cite{Qian2001}.}
  \label{fig3}
 \end{center}
\end{figure}

We then embarked on a determination of the in-plane ($k_x$-$k_y$)
and out-of-plane ($k_x$-$k_z$) FSs of the Fe films in order to trace the
electronic structure associated with the SS.
The in-plane (Fig.~\ref{fig3}(a)) and 
out-of-plane (Fig.~\ref{fig3}(b)) intensity maps  at $E_\mathrm{F}$
were obtained from 
polar angle scans, at $h\nu$ = 430~eV and as a function of energy 
$h\nu$ = 385--595~eV, respectively. 
The raw data was measured over more than one-quarter 
of the BZ (Fig.~\ref{fig3}(a) and (b)) \cite{EPAPS}.
From peaks in MDCs, we have determined the FS crossings 
and symmetrized to obtain the experimental in-plane and out-of-plane FSs 
for full BZ (Fig.~\ref{fig3}(c) and (d)). 
Since the fcc structure is equivalent along $\left<100\right>$, 
$\left<010\right>$ and $\left<001\right>$,
the in-plane and out-of-plane FSs should be equivalent in the absence of
electronic/magnetic instabilities along preferred directions.
The results instead show remarkable differences in the in-plane and
out-of-plane FSs. The in-plane FSs show a four-fold symmetry while the
out-of-plane FSs show two-fold symmetry. 
While the in-plane FSs show no evidence for nesting, a clear nesting is
observed in the out-of-plane FSs. This is directly seen in the MDC plot
comparison for the  $k_x$, $k_y$ and $k_z$ cuts along $\Gamma$-X
(Fig.~\ref{fig3}(e)). While the $k_x$ and $k_y$ cuts are very
similar, the $k_z$ cut shows one additional peak
between $\Gamma$ and X. 
Thus, the peak-to-peak separation for the symmetrized data,
marked by arrows in Fig.~\ref{fig3}(e), corresponds to
(2$\pi/a$)$\times$(0.86$\pm0.1$). The additional peaks in the MDCs are
observed 
over an extended region in $k_x$ (red dashed lines in
Fig.~\ref{fig3}(d)), thus identifying the  
$q_z$ component of the nesting vector. The identification of the SS
nesting vector in the out-of-plane FSs  is based on a precise
determination of reciprocal-space maps in successive BZs.  
In turn, this provides evidence for a subtle structural modification in
real space, namely, 
a small but finite compressive strain of 1.5$\pm$0.5~\% along the 
$c$-axis \cite{EPAPS}.

In order to determine the $q_x$ component of the SS vector, we
carried out an autocorrelation analysis (ACA)
\cite{Chatterjee2006,Shen2008, EPAPS}.
A full ACA over the BZ
did not lead to a meaningful result, and hence we resorted to a
restricted area ACA (red dashed lines in Fig.~\ref{fig3}(d)) 
so as to extract the $q_x$ component. The results are shown in
Fig.~\ref{fig3}(f) and (g), 
indicating that $q_x$ = 0.0 has the maximum spectral weight. Thus,
the experimental determination of FSs gives us a SS vector of
$\bm{q}$ = (2$\pi/a$)(0,0,0.86$\pm$0.1). While the MOKE results only provide
the $q_z$ component, since $q_x$ = 0.0 from the present results, it
indicates consistency between the MOKE and FS measurements within error
bars for the obtained SS vector (Fig.~\ref{fig3}(h)). 
 
For the epitaxial ultrathin Fe/Cu(001) films, 
all available band structure calculations are thus inconsistent
with the experimental results, as they do not give nested regions.
Also the magnitude of the SS does not match experiments.
While theoretical studies show that the SS in Erbium
is due to FS 
nesting, the SS in $\gamma$-Fe is not expected to be
derived from  
a single FS nesting vector \cite{Wessely2009}.
The above results indicate the requirement of an anisotropy between the 
$k_x$-$k_y$ and $k_x$-$k_z$, which probably originates in the
compressive strain along the $c$-axis \cite{EPAPS}.
Additionally, given that electron-electron correlations are not
negligible in Fe \cite{Hufner2000},
band structure calculations going beyond LSDA,
e.g.\ LSDA + DMFT with SS order, are considered necessary to explain the
observations. 
Finally, a recent theoretical study has shown that the SS in
fcc Fe is a candidate for spin transfer torque (STT: a current induced
transfer of vector spin between magnetic layers) \cite{Wessely2009}. 
STT was
predicted for the incommensurate $\bm{q}_1$ = (2$\pi/a$)(0,0,0.6),  but
it was shown to be suppressed for $\bm{q}_2$ = (2$\pi/a$)(0.5,0,1.0) due to
the antiferromagnetic coupling between layers. Our results indicate an
incommensurate SS vector of $\bm{q}$ = (2$\pi/a$)(0,0,$\sim$0.86),
suggesting the epitaxial Fe/Cu(001) thin film is a candidate for STT.  
Since STT gives rise to fascinating properties: 
microwave oscillations in a nanomagnet, current induced spin
waves \cite{Kiselev2003,Vlaminck2008},  
etc.\ 
and is technologically important for high-speed/high-density memory and
switching applications, 
it would be interesting to investigate the same in ultrathin epitaxial
Fe/Cu(001) films.
Furthermore, in-plane and out-of-plane FS mapping is expected to be
important for studying strain control of properties in ultrathin films
of elements, 
oxides, and their multilayers, heterostructures, interfaces, etc.

The present work has been performed under the approval of RIKEN
(Proposal No.~20080050).



\begin{thebibliography}{10}

\bibitem{Herring1966}
C. Herring, {\it Magnetism} (Academic Press, New York, 1966), Vol.~IV.

\bibitem{Wang1985}
C.~S. Wang, B.~M. Klein, and H. Krakauer, Phys. Rev. Lett. {\bf 54},  1852
  (1985).

\bibitem{Uhl1992}
M. Uhl, L.~M. Sandratskii, and J. K\"ubler, J. Magn. Magn. Mater. {\bf 103},
  314  (1992).

\bibitem{Korling1996}
M. K\"orling and J. Ergon, Phys. Rev. B {\bf 54},  R8293  (1996).

\bibitem{Okatov2009}
S. Okatov {\it et~al.}, Phys. Rev. B {\bf 79},  094111  (2009).

\bibitem{Shimizu2001}
K. Shimizu {\it et~al.}, Nature {\bf 412},  316  (2001).

\bibitem{Dubrovinsky1998}
L. Dubrovinsky {\it et~al.}, Science {\bf 281},  11a  (1998).

\bibitem{Tsunoda1989}
Y. Tsunoda, J. Phys.-Condes. Matter {\bf 1},  10427  (1989).

\bibitem{Tsunoda1993}
Y. Tsunoda, Y. Nishioka, and R.~M. Nicklow, J. Magn. Magn. Mater. {\bf 128},
  133  (1993).

\bibitem{Ramesh2007}
R. Ramesh and N. Spaldin, Nat. Mater. {\bf 6},  21  (2007).

\bibitem{Zhao2008}
J. Zhao {\it et~al.}, Nat. Mater. {\bf 7},  953  (2008).

\bibitem{Kief1993}
M.~T. Kief and W.~F. Egelhoff, Phys. Rev. B {\bf 47},  10785  (1993).

\bibitem{Ellerbrock1995}
R.~D. Ellerbrock {\it et~al.}, Phys. Rev. Lett. {\bf 74},  3053  (1995).

\bibitem{Bernhard2005}
T. Bernhard, M. Baron, M. Gruyters, and H. Winter, Phys. Rev. Lett. {\bf 95},
  087601  (2005).

\bibitem{Meyerheim2005}
H.~L. Meyerheim {\it et~al.}, Phys. Rev. B {\bf 71},  035409  (2005).

\bibitem{Li1994}
D.~Q. Li {\it et~al.}, Phys. Rev. Lett. {\bf 72},  3112  (1994).

\bibitem{Donath2009}
M. Donath, M. Pickel, A. Schmidt, and M. Weinelt, J. Phys.-Condes. Matter {\bf
  21},  134004  (2009).

\bibitem{Qian2001}
D. Qian {\it et~al.}, Phys. Rev. Lett. {\bf 87},  227204  (2001).

\bibitem{Amemiya2004}
K. Amemiya {\it et~al.}, Appl. Phys. Lett. {\bf 84},  936  (2004).

\bibitem{Spisak2002}
D. Spi\v{s}\'ak and J. Hafner, Phys. Rev. B {\bf 66},  052417  (2002).

\bibitem{Venturini2008}
F. Venturini {\it et~al.}, Phys. Rev. B {\bf 77},  045126  (2008).

\bibitem{Kamakura2006}
N. Kamakura {\it et~al.}, Phys. Rev. B {\bf 74},  045127  (2006).

\bibitem{Madjoe1999}
R.~H. Madjoe {\it et~al.}, J. Appl. Phys. {\bf 85},  6211  (1999).

\bibitem{Plucinski2008}
L. Plucinski {\it et~al.}, Phys. Rev. B {\bf 78},  035108  (2008).

\bibitem{Mansson2008}
M. M\r{a}nsson {\it et~al.}, Phys. Rev. Lett. {\bf 101},  226404  (2008).

\bibitem{EPAPS}
See EPAPS document No.~XXXXX for supplementary data.

\bibitem{Ohashi2007}
H. Ohashi {\it et~al.}, AIP Proc. {\bf 879},  523  (2007).

\bibitem{Takata2004}
Y. Takata {\it et~al.}, AIP Proc. {\bf 705},  1186  (2004).

\bibitem{Courths1984}
R. Courths and S. H\"ufner, Phys. Rep.-Rev. Sec. Phys. Lett. {\bf 112},  53
  (1984).

\bibitem{Mankey1993}
G.~J. Mankey, R.~F. Willis, and F.~J. Himpsel, Phys. Rev. B {\bf 48},  10284
  (1993).

\bibitem{Chatterjee2006}
U. Chatterjee {\it et~al.}, Phys. Rev. Lett. {\bf 96},  107006  (2006).

\bibitem{Shen2008}
D. Shen {\it et~al.}, Phys. Rev. Lett. {\bf 101},  226406  (2008).

\bibitem{Wessely2009}
O. Wessely, B. Skubic, and L. Nordstr\"om, Phys. Rev. B {\bf 79},  104433
  (2009).

\bibitem{Hufner2000}
S. H\"ufner {\it et~al.}, Phys. Rev. B {\bf 61},  12582  (2000).

\bibitem{Kiselev2003}
S. Kiselev {\it et~al.}, Nature {\bf 425},  380  (2003).

\bibitem{Vlaminck2008}
V. Vlaminck and M. Bailleul, Science {\bf 322},  410  (2008).

\end{thebibliography}
\end{document}